\documentclass[aps,preprint,amsmath]{revtex4}

\usepackage{graphicx}
\usepackage{slashed}
\usepackage{pifont}
\usepackage{tikz}
\usepackage{tkz-euclide}
\usepackage{makecell}

\def\al{\alpha}
\def\be{\beta}
\def\ga{\gamma}
\def\de{\delta}

\def\et{\eta}

\def\ka{\kappa}
\def\la{\lambda}

\def\om{\omega}

\def\cL{{\cal L}}

\def\mn{{\mu\nu}}

\def\frac#1#2{{\textstyle{{#1}\over{#2}}}}
\def\half{{\textstyle{1\over 2}}}

\def\prt{\partial}

\def\lsim{\mathrel{\rlap{\lower4pt\hbox{\hskip1pt$\sim$}}
    \raise1pt\hbox{$<$}}}
\def\gsim{\mathrel{\rlap{\lower4pt\hbox{\hskip1pt$\sim$}}
    \raise1pt\hbox{$>$}}}

\def\sqr#1#2{{\vcenter{\vbox{\hrule height.#2pt
         \hbox{\vrule width.#2pt height#1pt \kern#1pt
         \vrule width.#2pt}
         \hrule height.#2pt}}}}

\newcommand{\beq}{\begin{equation}}
\newcommand{\eeq}{\end{equation}}
\newcommand{\bea}{\begin{eqnarray}}
\newcommand{\eea}{\end{eqnarray}}
\newcommand{\rf}[1]{(\ref{#1})}

\newcommand{\bM}{\begin{pmatrix}}
\newcommand{\eM}{\end{pmatrix}}

\begin{document}

\title{Modifications to Plane Gravitational Waves from Minimal Lorentz Violation}

\author{Rui Xu}

\affiliation{Kavli Institute for Astronomy and Astrophysics, Peking University, Beijing {100871}, China}

\date{October 2019}

\begin{abstract}

{General Relativity predicts two modes for plane gravitational waves. When a tiny violation of Lorentz invariance occurs, the two gravitational wave modes are modified. We use perturbation theory to study the detailed form of the modifications to the two gravitational wave modes from the minimal Lorentz-violation coupling. The perturbation solution for the metric fluctuation up to the first order in Lorentz violation is discussed. Then, we investigate the motions of test particles under the influence of the plane gravitational waves with Lorentz violation. First-order deviations from the usual motions are found.}

\end{abstract}

\maketitle

\section{Introduction}
General Relativity (GR), as the standard classical gravitational theory, has been making predictions consistent with all the terrestrial experiments and most of the astrophysical observations \cite{review16, review162}. However, the fact that it is incompatible with quantum theory motivates ceaseless new tests and a large amount of alternative theories \cite{cmwreview, review}. Lorentz invariance, being one of the fundamental principles in GR, has~been suffering constant tests in various high-precision experiments and \mbox{observations \cite{datatable, llr07,ai08, shao14, shao16}}. Especially, gravitational wave observations, providing us the unique access to strong-field environments, {have}~recently put new stringent constraints on Lorentz violation based on the analysis of the modified dispersion relation of gravitational waves in the Standard-Model Extension (SME) framework \cite{km16, gw170817a}. 
  
The SME framework is a tool to study Lorentz violation in a model-independent way \cite{ck98, k06, km09,kt11,km12,km13, t14}. It incorporates all possible Lorentz-violation couplings into the Lagrangian density of GR and the Standard Model by employing the so called Lorentz-violation coefficients which can be measured or constrained with experimental data. The sector that describes gravity with Lorentz violation in vacuum is called the pure gravity sector of the SME \cite{bk06,bkx15}, and it is the theoretical basis from which the modified dispersion relation of gravitational waves is derived \cite{km16,km18}. 

Using the modified dispersion relation to constrain Lorentz violation marks the beginning of testing Lorentz invariance with gravitational wave observations \cite{cervantesgw16}. As the number and sensitivity of gravitational wave observatories increase \cite{kagra18, ligo19}, we can extract more information about the incoming waves from the observed signals, including the polarization status of them. Recently, a~detailed investigation on plane-wave solutions for arbitrary Lorentz violation in the pure gravitationa SME is carried out, and the modifications to the two polarization modes of the gravitational waves from coalescing compact binaries are considered \cite{m19}. Here we study a similar question but only with the simplest Lorentz-violation coupling in the pure gravity sector of the SME so that the calculations are more transparent.  We have to point out that there are much bigger indicators of Lorentz violation~\cite{sotiriou18}  {than what is described here.} Therefore our result is mostly pedagogical. In case it is to be used to constrain Lorentz violation in gravitational wave observations, a more comprehensive treatment to strain signals in gravitational-wave detectors is required.  

We start with describing the basics of the minimal Lorentz-violation coupling \cite{k06,bk06} and show that a plane wave ansatz gives a naive modification to the usual plane wave solution in Section \ref{2}. In~Section \ref{3}, we generalize the naive modification to serve as a rigorous perturbation solution to the Lorentz-violation field equations. In Section \ref{4}, the perturbation solution is used to find the geodesic deviation of test particles on a ring under the effect of gravitational waves with Lorentz violation.

\section{Plane Waves with Minimal Lorentz Violation}
\label{2}
The Lorentz-violation couplings in the SME framework are constructed as coordinate scalars of the Lorentz-violation coefficients and conventional field operators. The simplest term in the pure gravity sector is \cite{k06,bk06}
\bea
\cL^{(4)} = \frac{1}{16\pi G} ( -u R + s^\mn R^{T}_\mn + t^{\al\be\ga\de} C _{\al\be\ga\de} ) ,
\label{mgSME}
\eea  
where $u, s^\mn$, and $t^{\al\be\ga\de}$ are called the minimal Lorentz-violation coefficients as the coupling involves no derivatives of the Riemann tensor. $s^\mn$ and $t^{\al\be\ga\de}$ inherit the symetries and traceless property of the trace-free Ricci tensor, $R^{T}_\mn$, and the Weyl conformal tensor, $C_{\al\be\ga\de}$, separately. Note that the superscript $4$ on $\cL$ represents the mass dimension of the gravitational operators (including the gravitational constant factor $G$). Therefore, the minimal Lorentz-violation coefficients $u, s^\mn$, and $t^{\al\be\ga\de}$, are also called the Lorentz-violation coefficients with mass dimension $d=4$.

Adding to the Einstein--Hilbert term, the Lagrangian density \rf{mgSME} gives modifications from minimal Lorentz-violation to the Einstein field equations. The details on linearizing the modified field equations and expressing them in terms of the background values of $u, s^\mn$, and $t^{\al\be\ga\de}$ are demonstrated in Ref.~\cite{bk06}. Here we just show the result which is the starting point of our calculation, namely the linearized vacuum field equations with minimal Lorentz-violation. They are 
\bea
R_\mn = \bar s^{\al\be} R_{\al\mn\be} ,
\label{mgeom}
\eea
with $\bar s^{\al\be}$ being the background value of $s^\mn$. Note that the background values of $u$ and $t^{\al\be\ga\de}$ do not appear \cite{bk06}. We also point out that the word "linearized" has two meanings here. One is the same as usual, namely the gravitational field is linearized. The second is that Equation \rf{mgeom} holds up to the first order in $\bar s^{\al\be}$. There is no need to keep terms at higher orders in $\bar s^{\al\be}$ because Lorentz violation should be tiny to be consistent with the experimental support for Lorentz invariance.  

The dispersion relation implied by a generalized form of Equation \rf{mgeom} is studied in Ref. \cite{kt15} to predict gravitational \v{C}erenkov radiation from Lorentz violation. They proposed the modified harmonic gauge condition,
\bea
(\et^{\la\ka} + \bar s^{\la\ka} ) \prt_\la h_{\ka\mu} = \half ( \et^{\la\ka} + \bar s^{\la\ka} ) \prt_\mu h_{\la\ka} , 
\label{gauge}
\eea
that simplifies the field equations \rf{mgeom} to
\bea
(\et^{\al\be}  + {\bar s}^{\al\be} ) \prt_\al \prt_\be h_{\mn} = 0 ,
\label{mgeom2}
\eea
where $h_\mn = g_\mn - \et_\mn $ is the fluctuation of the metric.
Using the plane wave ansatz 
\bea
h_\mn(x) = A_\mn e^{ikx} ,
\label{mpw}
\eea
the modified dispersion relation up to the first order in $\bar s^{\al\be}$ is found to be
\bea
k^0 = |\vec k| + \half \frac{\bar s^{\al\be}  k_\al k_\be}{|\vec k|} .
\label{dp}
\eea
Namely the wave vector can be written as $k^\mu = (\om + \de \om , \,  \vec k)$ with $\om = |\vec k|$ and $\de \om = \half \frac{\bar s^{\al\be}  k_\al k_\be}{|\vec k|}$. Thus, the plane wave solution can be written as
\bea
h_\mn(x) = A_\mn e^{-i(\om t- \vec k \cdot \vec x)} - i \, \de \om \, t \, A_\mn e^{-i(\om t- \vec k \cdot \vec x)} + ... .
\label{mpwsol}
\eea

The first term, $A_\mn e^{-i(\om t- \vec k \cdot x)}$, is apparently the plane wave solution in GR, and the rest consists of corrections from Lorentz violation. Up to the first order in $\bar s^{\al\be}$, the correction is 
\bea
h^{(1)}_\mn = - i \, \de \om \, t \, A_\mn e^{-i(\om t- \vec k \cdot \vec x)} .
\label{mpwh1}
\eea
{As there is a factor of $t$ in the amplitude of $h^{(1)}_\mn$, the correction is only valid during a finite time period. The~plane wave solution \rf{mpwsol} is insufficient to describe the entire content of the Lorentz-violation modification to gravitational waves. However, Equation \rf{mpwh1} provides us an insight into how the modification might look. In the next section, we will take the generalized form of Equation \rf{mpwh1}, which~is}
\bea
h^{(1)}_\mn = C_{\mn\al} x^\al e^{-i(\om t- \vec k \cdot \vec x)} ,
\label{h1}
\eea
as an ansatz to solve the field equations \rf{mgeom2} up to the first order in $\bar s^{\al\be}$. The constants $C_{\mn\al}$ are going to be determined by the gauge condition \rf{gauge} and the field equations \rf{mgeom2}. Note that the Lorentz-violation modification shown in \rf{h1} applies only to a finite spacetime region as the coordinates $x^\al$ appear in the amplitude.

\section{The Perturbation Solution}
\label{3}
We seek a perturbation solution up to the first order in $\bar s^{\al\be}$ for the field equations \rf{mgeom2}. To proceed, we assume that the zeroth-order plane wave travels along the $z$ direction with the conventional wave vector 
\bea
k^{(0)\mu} = (\om,\,  0,\,  0,\,  \om) ,
\label{0wv}
\eea
and that its amplitude $A_\mn$ takes the usual form
\bea
A_\mn = 
\begin{pmatrix}
0 & 0 & 0 & 0\\
0 & A_{11} & A_{12} & 0\\
0 & A_{12} & -A_{11} & 0\\
0 & 0 & 0 & 0 
\end{pmatrix} ,
\label{0thA}
\eea  
where $A_{11}$ is the amplitude of the plus wave and $A_{12}$ is the amplitude of the cross wave. By substituting 
\bea
h_\mn(x) = A_\mn e^{-i(\om t-  k z)} + C_{\mn\al} x^\al e^{-i(\om t-  k z)}  ,
\label{hsol}
\eea
into the field equations \rf{mgeom2} and keeping only the first-order terms, we have 
\bea
 2 C_{\mn\al} i k^{(0) \al}  = \bar s^{\al\be} k^{(0)}_\al k^{(0)}_\be A_{\mn} .
\eea
Writting the above equations explicitly, they are
\bea
C_{\mn 0} + C_{\mn 3} = -\frac{i\om }{2} (\bar s^{00} - 2 \bar s^{03} + \bar s^{33} ) A_{\mn} .
\label{c1}
\eea
In addition, using Equation~\rf{hsol} in the gauge condition \rf{gauge}, up to the first order we have
\bea
&& \et^{\la\ka}  C_{\ka\mu\al} i k^{(0)}_\la x^\al + \et^{\la\ka}  C_{\ka\mu\la} + \bar s^{\la\ka} i k^{(0)}_\la A_{\ka\mu} =  \half  \left( C_\al  i k^{(0)}_\mu x^\al + C_\mu + \bar s^{\la\ka} i k^{(0)}_\mu A_{\la\ka} \right)  ,
\label{eq1}
\eea
where $C_\al = \et^{\mn} C_{\mn\al}$. The relations \rf{eq1} imply two sets of equations:
\bea
&& \et^{\la\ka}  C_{\ka\mu\al} k^{(0)}_\la   - \half  C_\al  k^{(0)}_\mu = 0  ,
\label{cc2}
\eea
and
\bea
&& \et^{\la\ka}  C_{\ka\mu\la} - \half  C_\mu =   \frac{i}{2} \bar s^{\la\ka} k^{(0)}_\mu A_{\la\ka} - i \bar s^{\la\ka} k^{(0)}_\la A_{\ka\mu} .
\label{cc3}
\eea
Using the expressions \rf{0wv} and \rf{0thA}, we find that Equation \rf{cc2} can be simplified to
\bea
  C_{00\al} + 2 C_{03\al} + C_{33\al}  &=& 0 ,
\nonumber \\
  C_{11\al} + C_{22\al}  &=& 0 ,
 \\
 C_{01\al} + C_{31\al} &=& 0 ,
\nonumber \\
 C_{02\al} + C_{32\al} &=& 0 , \nonumber
\label{c2}
\eea
and that Equation~\rf{cc3} can be simplified to
\bea
  C_{011} + C_{022}  &=& - \half i\om \big( (\bar s^{11} - \bar s^{22} ) A_{11} + 2 \bar s^{12} A_{12} \big) ,
\nonumber \\
  C_{111} + C_{122} + \half (C_{001}-C_{331} )  &=& i\om A_{11} (\bar s^{01} - \bar s^{31})  +  i\om A_{12} ( \bar s^{02} - \bar s^{32} ) ,
\\
  C_{121} - C_{112} + \half (C_{002}-C_{332} )  &=& i\om A_{12} (\bar s^{01} - \bar s^{31})  -  i\om A_{11} ( \bar s^{02} - \bar s^{32} ) .\nonumber 
\label{c3}
\eea
Note that Equation~\rf{cc3} turns out to have only $3$ independent equations.

Equation \rf{c2} shows that there are $6$ independent components in the first-order solution $h^{(1)}_\mn$, which can be written as
\bea
h^{(1)}_\mn = 
\begin{pmatrix}
h^{(1)}_{00} & h^{(1)}_{01} & h^{(1)}_{02} & -\half (h^{(1)}_{00} + h^{(1)}_{33})\\
h^{(1)}_{01} & h^{(1)}_{11} & h^{(1)}_{12} & -h^{(1)}_{01}\\
h^{(1)}_{02} & h^{(1)}_{12} & -h^{(1)}_{11} & -h^{(1)}_{02}\\
-\half (h^{(1)}_{00} + h^{(1)}_{33}) & -h^{(1)}_{01} & -h^{(1)}_{02} & h^{(1)}_{33} 
\end{pmatrix} .
\eea
The $6$ independent components are easily divided into 3 groups, $\{ h^{(1)}_{11},\,  h^{(1)}_{12} \} $, $\{ h^{(1)}_{00}, \, h^{(1)}_{33} \}$, and $\{ h^{(1)}_{01},\,  h^{(1)}_{02} \}$. The remaining equations in \rf{c1} and \rf{c3} are insufficient to determine any of them. This indicates that the ansatz \rf{h1} does not lead to a unique first-order solution. We need extra information to fix $h^{(1)}_\mn$. Next, we discuss the solutions for $\{ h^{(1)}_{11},\,  h^{(1)}_{12} \} $, $\{ h^{(1)}_{00}, \, h^{(1)}_{33} \}$ and $\{ h^{(1)}_{01}, \, h^{(1)}_{02} \}$ separately.  

\subsection{$ \{ h^{(1)}_{11}, \, h^{(1)}_{12} \}$ }
We expect these two components recover the correction \rf{mpwh1}. This is indeed the case if we take all the components of $C_{11\al}$ and  $C_{12\al}$ to be zero except for
\bea
 C_{11 0} &=& -\frac{i\om }{2} (\bar s^{00} - 2 \bar s^{03} + \bar s^{33} ) A_{11},
\nonumber \\
 C_{12 0} &=& -\frac{i\om }{2} (\bar s^{00} - 2 \bar s^{03} + \bar s^{33} ) A_{12}.
\eea  
In this way, $h^{(1)}_{11}$ and $h^{(1)}_{12}$ are fixed, and the dispersion relation \rf{dp} can be recovered in the perturbation~solution.  

\subsection{$ \{ h^{(1)}_{00}, \, h^{(1)}_{33} \}$ }
With $C_{111} = C_{112} = C_{121} = C_{122} = 0$, we have 
\bea
  C_{001}-C_{331}  &=& 2 i\om A_{11} (\bar s^{01} - \bar s^{31})  +  2 i\om A_{12} ( \bar s^{02} - \bar s^{32} ) ,
\nonumber \\
  C_{002}-C_{332}   &=& 2 i\om A_{12} (\bar s^{01} - \bar s^{31})  -  2 i\om A_{11} ( \bar s^{02} - \bar s^{32} ) .
\label{h00h33}
\eea
It turns out the combinations $C_{001}-C_{331}$ and $C_{002}-C_{332}$ are the only terms involving $C_{00\al}$ and $C_{33\al}$ in the first-order Riemann tensor (see the Appendix \ref{a}). Therefore, without any ambiguity in observables, we can safely assume all the components of $C_{00\al}$ and $C_{33\al}$ vanishing except for $C_{001}$ and $C_{002}$, which are given by Equation \rf{h00h33}.

\subsection{$ \{ h^{(1)}_{01}, \, h^{(1)}_{02} \}$ }
In the Appendix \ref{a}, we can see that $C_{010},\ , C_{013} ,\, C_{020}$, and $C_{023}$ do not appear in the first-order Riemann tensor. Therefore, they can be taken as zero. However, $C_{011}$ and $C_{022}$ appear, and they do not appear as the combination $C_{011} +  C_{022}$ as shown in Equation \rf{c3}. In addition, $C_{012}$ and $C_{021}$ also show up in the first-order Riemann tensor. Namely, we have one equation in \rf{c3} to use but $4$ unknowns, $C_{011}, \, C_{012},\, C_{021}$, and $C_{022}$, to determine. The inadequacy is likely from the fact that we are missing certain information about the specific dynamic model of the Lorentz-violation coefficient $s^{\al\be}$. In other words, we expect $s^{\al\be}$ to have its own field equations with the metric involved. Then, when $s^{\al\be}$ is approximated by its background value $\bar s^{\al\be}$, some of these field equations degenerate to constraints on the metric though most of them vanish trivially.

Building a specific dynamic model for $s^{\al\be}$ simply lies beyond the scope of the present work. For~the calculation in the next section, we decide to choose the simplest solution for $h^{(1)}_{01}$ and $h^{(1)}_{02}$, by which we mean that all the components of $C_{01\al}$ and $C_{02\al}$ vanish except for
\bea
C_{011}  = - \half i\om \big( (\bar s^{11} - \bar s^{22} ) A_{11} + 2 \bar s^{12} A_{12} \big) . 
\eea
\section{Geodesic Deviation}
\label{4}
Now we use the above first-order solution to calculate the effects of Lorentz violation on the motions of test particles when plane gravitational waves pass through. Similarly to the usual case, it is illustrative to consider a ring of test particles whose initial positions form a circle 
\bea
( X(0) )^2 + ( Y(0) )^2 = d^2 ,
\eea
in a local inertial frame with local coordinates $ \{X,\,Y, \,Z \} $. Assuming the local coordinates are aligned with the general coordinates $\{ x,\,y,\,z \}$, then the nonrelativistic geodesic deviation equations that determine the motions of the test particles in the local frame are \cite{b2}
\bea
\frac{d^2 X}{dt^2} &=&  - R_{0 1 01} X(0) - R_{0 1 02} Y(0) - R_{0 1 03} Z(0) , 
\nonumber \\
\frac{d^2 Y}{dt^2} &=&  - R_{0 2 01} X(0) - R_{0 2 02} Y(0) - R_{0 2 03} Z(0)  ,
\\
\frac{d^2 Z}{dt^2} &=&  - R_{0 3 01} X(0) - R_{0 3 02} Y(0) - R_{0 3 03} Z(0)  .\nonumber 
\eea  
The zeroth-order solution for $X(t), \,Y(t)$ and $Z(t)$ is the usual deformation  
\bea
 X^{(0)} (t) - X(0) &=&  \half \big( A_{11} X(0) + A_{12} Y(0) \big) ( e^{-i\om t} - 1) ,
\nonumber \\
 Y^{(0)} (t) - Y(0) &=&  \half \big( A_{12} X(0) - A_{11} Y(0) \big) ( e^{-i\om t} - 1),
 \\
 Z^{(0)} (t) - Z(0) &=&  0 .\nonumber
\label{0th}
\eea
Note that we have assumed that the local frame is moving along the geodesic $x(t) = y(t) = z(t) = 0$. The first-order solution turns out to be
\bea
 X^{(1)}(t) &=&  -\frac{1}{\om^2} \big( \al_X + \be_X( \om t - 2i) \big) e^{-i\om t} + \frac{1}{\om^2} ( \al_X - 2i \be_X) ,
\nonumber \\
 Y^{(1)}(t) &=&  -\frac{1}{\om^2} \big( \al_Y + \be_Y( \om t - 2i) \big) e^{-i\om t} + \frac{1}{\om^2} ( \al_Y - 2i \be_Y)   ,
\\
 Z^{(1)}(t) &=&  -\frac{1}{\om^2}  \al_Z  e^{-i\om t} + \frac{1}{\om^2}  \al_Z ,\nonumber 
\label{1st}
\eea
where
\bea
 \al_X &=&   - i\om ( C_{110} - C_{011} ) X(0) -  i\om C_{120}   Y(0)  + \frac{1}{4} i\om C_{001} Z(0) ,
\nonumber \\
 \be_X &=&  - \half  \om C_{110}  X(0) - \half  \om C_{120}  Y(0)  ,
\nonumber \\
 \al_Y &=&   - i\om C_{120}  X(0) + i\om C_{110} Y(0)  + \frac{1}{4} i\om C_{002} Z(0) ,
\\
 \be_Y &=&  - \half  \om C_{120}  X(0) + \half  \om C_{110}  Y(0)  ,
\nonumber \\
 \al_Z &=&  \frac{1}{4} i\om  C_{001} X(0) + \frac{1}{4} i\om  C_{002} Y(0)  .  \nonumber 
\eea
The solution \rf{0th} as well as Equation \rf{1st} is written with the understanding that only the real parts are taken.

The most notable correction is that Lorentz violation causes an oscillation along the $z$ direction in general, which does not happen in the case of the usual plane gravitational waves. Then, for the ring of the test particles in the $XY$ plane, we find that the shape is still deformed into ellipses. But the semi axes are corrected by Lorentz violation. Specifically speaking, when $A_{11}$ is real and $A_{12} = 0$, the semi axes of the ellipse at time $t$ are
\bea
 a &=& d \big( 1 + \half A_{11} (\cos{\om t} -1) - \half A_{11} (\bar s^{11} - \bar s^{22} )(\cos{\om t} -1) - \frac{1}{4} A_{11} (\bar s^{00} - 2 \bar s^{03} + \bar s^{33}) \om t \sin{\om t} \big) ,
\nonumber \\
 b &=& d \big( 1 - \half  A_{11} (\cos{\om t} -1) + \frac{1}{4} A_{11} (\bar s^{00} - 2 \bar s^{03} + \bar s^{33}) \om t \sin{\om t}  \big) ;
\eea
when $A_{12}$ is real and $A_{11} = 0$, the semi axes of the ellipse at time $t$ are
\bea
 a &=& d \big( 1 + A_{12} (\cos{\om t} -1) - A_{12} \bar s^{12} (\cos{\om t} -1) - \half A_{12} (\bar s^{00} - 2 \bar s^{03} + \bar s^{33}) \om t \sin{\om t} \big)  ,
\nonumber \\
 b &=& d \big( 1 - A_{12} (\cos{\om t} -1) - A_{12} \bar s^{12} (\cos{\om t} -1) + \half A_{12} (\bar s^{00} - 2 \bar s^{03} + \bar s^{33}) \om t \sin{\om t}  \big) .
\eea
Last but not least, we point out that when $A_{12}$ is real and $A_{11} = 0$, the rotation angle of the ellipses from the standard position 
\bea
\frac{X^2}{a^2} + \frac{Y^2}{b^2} = 1 ,
\eea
is not $\pm \frac{\pi}{4}$ any more. A time-independent deviation of $\half \bar s^{12}$ occurs in the presence of Lorentz violation.  
\section{Conclusions}

We used the ansatz \rf{h1} to find the correction to plane gravitational waves from minimal Lorentz violation. It was shown that up to the first order in Lorentz violation, the { correction}, $h^{(1)}_\mn$, {has} $6$ independent components, with $4$ of them fixed in the SME framework. To determine the remaining two components, extra information about the dynamics of the Lorentz-violation coefficient $s^{\al\be}$ is necessary. This requires treating $s^{\al\be}$ as a dynamic field and assigning it a kinetic term in the Lagrangian density. {This} lies beyond the scope of the present work. 

Then, to demonstrate the effects of Lorentz violation on the motions of test particles under the influence of plane gravitational waves, we artificially fixed the two undetermined components of $h^{(1)}_\mn$. Together with the other 4 determined components, two notable effects were found. One is the oscillation of a test particle along the propagating direction of the gravitational waves, and the other is the deviation from $\pm \frac{\pi}{4}$ for the rotation angle of the deformed ellipse in the presence of the cross wave. 
  {Note that the amplitude of the oscillation along the $Z$-direction is proportional to the amplitude of the zeroth-order gravitational wave but suppressed by the components of the Lorentz-violation coefficient $\bar s^{\al\be}$. Taking the current best bound of $10^{-15}$} \cite{gw170817a}  {on $\bar s^{\al\be}$ into consideration, it is unlikely that this oscillation provides a viable test of Lorentz violation even in the near future. On the other hand, as we are getting access to the polarization information of incoming gravitational waves with more and more detectors in construction, the deviation of the rotation angle suggests a Lorentz-violation phase difference between the two polarization modes to test in future observations of polarized gravitational waves. To conduct such tests, a more comprehensive treatment in the context of existing and future gravitational-wave detectors is required, which deserves another paper for investigation.}

\section*{Acknowledgments}

R.X. is thankful to  Alan Kosteleck\'{y},  Lijing Shao, and  Jay Tasson for valuable comments.

\appendix
\section{The First-Order Riemann Tensor}
\label{a}
The first-order Riemann tensor is calculated by 
\bea
R^{(1)}_{\al\be\ga\de} = \half (\prt_\ga \prt_\be h^{(1)}_{\al\de} + \prt_\al \prt_\de h^{(1)}_{\be\ga} - \prt_\ga \prt_\al h^{(1)}_{\be\de} - \prt_\de \prt_\be h^{(1)}_{\al\ga}) .
\eea
Plugging Equation \rf{h1} into it, and using $\vec k = (0,\,0,\,\om)$, we find
\bea
 R^{(1)}_{0101} &=&  \half \big( 2 i\om ( C_{110} - C_{011} ) + \om^2 C_{11\al} x^\al \big) e^{-i(\om t- \vec k \cdot \vec x)} ,
\nonumber \\
 R^{(1)}_{0102} &=&  \half \big( i\om ( 2C_{120} - C_{021} - C_{012} ) + \om^2 C_{12\al} x^\al \big) e^{-i(\om t- \vec k \cdot \vec x)} ,
\nonumber \\
 R^{(1)}_{0103} &=& -\frac{1}{4} i\om (C_{001} -C_{331})  e^{-i(\om t- \vec k \cdot \vec x)} ,
\\
 R^{(1)}_{0202} &=& -\half \big( 2i\om (  C_{110} + C_{022} ) + \om^2 C_{11\al} x^\al \big) e^{-i(\om t- \vec k \cdot \vec x)} ,
\nonumber  \\
 R^{(1)}_{0203} &=& -\frac{1}{4} i\om (C_{002} -C_{332})  e^{-i(\om t- \vec k \cdot \vec x)} ,
\nonumber  \\
 R^{(1)}_{0303} &=& 0 ,\nonumber
\eea \vspace{-20pt}
\bea
 R^{(1)}_{0112} &=& - \half i\om ( C_{112} - C_{121} ) e^{-i(\om t- \vec k \cdot \vec x)} ,
\nonumber \\
 R^{(1)}_{0113} &=& \half \big( i\om ( C_{110} - C_{113} - 2 C_{011} ) + \om^2 C_{11\al} x^\al  \big) e^{-i(\om t- \vec k \cdot \vec x)} ,
\nonumber \\
 R^{(1)}_{0123} &=& -\half  i\om C_{012} e^{-i(\om t- \vec k \cdot \vec x)} ,
\nonumber \\
 R^{(1)}_{0212} &=& - \half  i\om ( C_{111} + C_{122} ) e^{-i(\om t- \vec k \cdot \vec x)} ,
\nonumber \\
 R^{(1)}_{0213} &=& \half \big( i\om ( C_{120} - C_{123} - C_{021} - C_{012} ) + \om^2 C_{12\al} x^\al  \big) e^{-i(\om t- \vec k \cdot \vec x)} ,
\\
 R^{(1)}_{0223} &=& \half \big( - i\om ( C_{110} - C_{113} + 2 C_{022} ) - \om^2 C_{11\al} x^\al ) e^{-i(\om t- \vec k \cdot \vec x)} ,
\nonumber \\
 R^{(1)}_{0313} &=& - \frac{1}{4} i\om ( C_{001} - C_{331} ) e^{-i(\om t- \vec k \cdot \vec x)} ,
\nonumber \\
 R^{(1)}_{0323} &=& - \frac{1}{4} i\om ( C_{002} - C_{332} ) e^{-i(\om t- \vec k \cdot \vec x)} ,\nonumber  
\eea
and
\bea
 R^{(1)}_{1212} &=& 0 ,
\nonumber \\
 R^{(1)}_{1213} &=& - \half  i\om  ( C_{112} - C_{121} ) e^{-i(\om t- \vec k \cdot \vec x)} ,
\nonumber \\
 R^{(1)}_{1223} &=& - \half  i\om ( C_{111} + C_{122} ) e^{-i(\om t- \vec k \cdot \vec x)} ,
\nonumber \\
 R^{(1)}_{1313} &=&  \half  \big( -2 i\om ( C_{113} + C_{011} ) + \om^2 C_{11\al} x^\al \big) e^{-i(\om t- \vec k \cdot \vec x)} ,
\\
 R^{(1)}_{1323} &=& \half \big( - i\om ( C_{012} + C_{021} + 2 C_{123} ) + \om^2 C_{12\al} x^\al \big) e^{-i(\om t- \vec k \cdot \vec x)} ,
\nonumber  \\
 R^{(1)}_{2323} &=& \half \big( 2i\om ( C_{113} - C_{022} ) - \om^2 C_{11\al} x^\al \big) e^{-i(\om t- \vec k \cdot \vec x)}.\nonumber 
\eea


\begin{thebibliography}{99}

\bibitem[Ivan(2016)]{review16}
Debono, I.; Smoot, G.F. General Relativity and Cosmology: Unsolved Questions and Future Directions. {\it Universe} {\bf 2016}, \emph{2}, 23.

\bibitem[Vishwakarma(2016)]{review162}
Vishwakarma, R.G. Einstein and Beyond: A Critical Perspective on General Relativity. {\it Universe} {\bf 2016}, \emph{2}, 11.

\bibitem[Will(2014)]{cmwreview}
Will, C.M. The confrontation between general relativity and experiment. {\it Living Rev. Relativ.} {\bf 2014}, \emph{14}, 4.

\bibitem[Will(2015)]{review}
Berti, E.; Barausse, E.; Cardoso, V.; Gualtieri, L.; Pani, P.; Sperhake, U.; Stein, L.C.; Wex, N.; Yagi, K.; \mbox{Baker, T.; et al.} Testing General Relativity with Present and Future Astrophysical Observations. {\it Class. Quantum Grav.} {\bf 2015}, \emph{32}, 243001.

\bibitem[NA(2011)]{datatable}
Kosteleck\'{y}, V.A.; Russell, N. Data tables for Lorentz and CPT violation. {\it Rev. Mod. Phys.} {\bf 2011}, \emph{83}, 11.

\bibitem[llr(2007)]{llr07}
Battat, J.B.R.; Chandler, J.F.; Stubbs, C.W. Testing for Lorentz Violation: Constraints on Standard-Model Extension Parameters via Lunar Laser Ranging. {\it Phys. Rev. Lett.} {\bf 2007}, \emph{99}, 241103.

\bibitem[ai(2008)]{ai08}
Muller, H.; Chiow, S.-W.; Herrmann, S.; Chu, S.; Chung, K.-Y. Atom Interferometry tests of the isotropy of post-Newtonian gravity. {\it Phys. Rev. Lett.} {\bf 2008}, \emph{100}, 031101.

\bibitem[shao(2014)]{shao14}
Shao, L. Tests of local Lorentz invariance violation of gravity in the standard model extension with pulsars. {\it Phys. Rev. Lett.} {\bf 2014}, \emph{112}, 111103.

\bibitem[shao(2016)]{shao16}
Shao, C.-G.; Tan, Y.-J.; Tan, W.-H.; Yang, S.-Q.; Luo, J.; Tobar, M.E.; Bailey, Q.G.; Long, J.C.; Weisman, E.;
\mbox{Xu, R.; et al.} Combined search for Lorentz violation in short-range gravity. {\it Phys. Rev. Lett.} {\bf 2016}, \emph{117}, 071102.



\bibitem[km(2016)]{km16}
Kosteleck\'{y}, V.A.; Mewes, M. Testing local Lorentz invariance with gravitational waves. {\it Phys. Lett. B} {\bf 2016}, \emph{757}, 510.

\bibitem[abbott(2017)]{gw170817a}
Abbott, B.P.; Abbott, R.; Abbott, T.D.; Acernese, R.; Ackley, K.; Adams, C.; Adams, T.; Addesso, P.; Adhikari,~R.X.; Adya, V.B.; et al. Gravitational waves and gamma-rays from a binary Neutron star merger: GW170817
and GRB 170817A. {\it Astrophys. J. Lett.} {\bf 2017}, \emph{848}, L13.

\bibitem[ck(1998)]{ck98}
Colladay, D.; Kosteleck\'{y}, V.A. Lorentz-violating extension of the Standard Model. {\it Phys. Rev. D} {\bf 1998}, \emph{58}, 116002.

\bibitem[k(2004)]{k06}
Kosteleck\'{y}, V.A. Gravity, Lorentz violation, and the Standard Model. {\it Phys. Rev. D} {\bf 2004}, \emph{69}, 105009.

\bibitem[km(2009)]{km09}
Kosteleck\'y, V.A.; Mewes, M. Electrodynamics with Lorentz-violating operators of arbitrary dimension.
{\it \mbox{Phys. Rev. D}} {\bf 2009}, \emph{80}, 015020.

\bibitem[kt(2011)]{kt11}
Kosteleck\'{y}, V.A.; Tasson, J.D. Matter-gravity couplings and Lorentz violation. {\it Phys. Rev. D} {\bf 2011}, \emph{83}, 016013.

\bibitem[km(2012)]{km12}
Kosteleck\'{y}, V.A.; Mewes, M. Neutrinos with Lorentz-violating operators of arbitrary dimension. {\it \mbox{Phys. Rev. D}} {\bf 2012}, \emph{85}, 096005.

\bibitem[km(2013)]{km13}
Kosteleck\'{y}, V.A.; Mewes, M. Fermions with Lorentz-violating operators of arbitrary dimension. {\it Phys. Rev. D} {\bf 2013}, \emph{88}, 096006.

\bibitem[t(2014)]{t14}
Tasson, J.D. What do we know about Lorentz invariance? {\it Rep. Prog. Phys.} {\bf 2014}, \emph{77}, 062901.



\bibitem[bk(2006)]{bk06}
Bailey, Q.G.; Kosteleck\'{y}, V.A. Signals for Lorentz violation in post-newtonian gravity. {\it Phys. Rev. D} {\bf 2006}, \emph{74}, 045001.

\bibitem[bkx(2015)]{bkx15}
Bailey, Q.G.; Kosteleck\'{y}, V.A.; Xu, R. Short-range gravity and Lorentz violation. {\it Phys. Rev. D} {\bf 2015}, \emph{91}, 022006.

\bibitem[km(2018)]{km18}
Kosteleck\'{y}, V.A.; Mewes, M. Lorentz and Diffeomorphism Violations in Linearized Gravity. {\it Phys. Lett. B} {\bf 2018}, \emph{779}, 136.

\bibitem[Cervantes(2016)]{cervantesgw16}
Cervantes-Cota, J.L.; Galindo-Uribarri, S.; Smoot, G.F. A Brief History of Gravitational Waves. {\it Universe} {\bf 2016}, \emph{2}, 22.

\bibitem[kagra(2018)]{kagra18}
Akutsu, T.; Ando, M.; Arai, K.; Arai, Y.; Araki, S.; 
Araya, A.; Aritomi, N.; Asada, H.; Aso, Y.;  Atsuta, S.; et al. KAGRA:
2.5 Generation Interferometric Gravitational Wave
Detector. \emph{arXiv} {\bf 2018}, arXiv:1811.08079.

\bibitem[ligo(2019)]{ligo19}
Abbott, B.P.; Abbott, R.; Abbott, T.D.; Abernathy, M.R.; Acernese, F.; Ackley, K.; Adams, C.; Adams, T.;
Addesso, P.; Adhikari, R.X.; et al. LIGO scientific collaboration and virgo collaboration. Gravitational wave
astronomy with LIGO and similar detectors in the
next decade. \emph{arXiv} {\bf 2019}, arXiv:1904.03187.

\bibitem[m(2019)]{m19}
Mewes, M. Signals for Lorentz violation in gravitational waves. {\it Phys. Rev. D} {\bf 2019}, \emph{99}, 104062.

\bibitem[Sotiriou(2018)]{sotiriou18}
Sotiriou, T.P. Detecting Lorentz Violations with Gravitational Waves from Black Hole Binaries. {\it \mbox{Phys. Rev. Lett.}} {\bf 2018}, \emph{120}, 041104.

\bibitem[kt(2015)]{kt15}
Kosteleck\'{y}, V.A.; Tasson, J.D. Constraints on Lorentz violation from gravitational \v{C}erenkov radiation. {\it \mbox{Phys. Lett. B}} {\bf 2015}, \emph{749}, 551.
\bibitem[Will(2014)]{b2}
Poisson E.; Will, C.M. {\it Gravity}; Cambridge University Press: Cambridge, UK, {2014}.
\end{thebibliography}
\end{document}